\documentclass{svjour3}                     
\smartqed  
\usepackage{graphicx}
\usepackage{mathptmx}      
\usepackage{natbib}
\newcommand{\beq}{\begin{equation}}
\newcommand{\eeq}{\end{equation}}
\newcommand{\beqa}{\begin{eqnarray}}
\newcommand{\eeqa}{\end{eqnarray}}

\newcommand{\um}{$\mu$m}
\newcommand{\Angstrom}{\,{\rm\AA}}
\newcommand{\AU}{\,{\rm AU}}
\newcommand{\cm}{\,{\rm cm}}
\newcommand{\gram}{{\,{\rm g}}}
\newcommand{\gtsim}{\lower.5ex\hbox{$\; \buildrel > \over \sim \;$}} 
\newcommand{\K}{\,{\rm K}}
\newcommand{\kms}{\,{\rm km~s}^{-1}}
\newcommand{\kpc}{\,{\rm kpc}}
\newcommand{\ltsim}{\lower.5ex\hbox{$\; \buildrel < \over \sim \;$}}
\newcommand{\micron}{\,\mu{\rm m}}
\newcommand{\Msol}{{M_\odot}}
\newcommand{\Myr}{{\rm Myr}}
\newcommand{\pc}{\,{\rm pc}}
\newcommand{\yr}{\,{\rm yr}}
\newcommand{\newtext}[1]{#1}
\journalname{SSRv}
\begin{document}
\title{Perspectives on Interstellar Dust Inside and Outside of the Heliosphere
}


\titlerunning{Perspectives on Interstellar Dust in the Heliosphere}

\author{B. T. Draine}


\institute{Princeton University Observatory \at
              Peyton Hall\\
              Princeton University\\
	      Princeton, NJ 08544-1001 USA\\
              Tel.: +01-609-258-3810\\
              Fax: +01-609-258-1020\\
              \email{draine@astro.princeton.edu}           
}

\date{Received: date / Accepted: date}

\maketitle
\begin{abstract}
Measurements by dust detectors on interplanetary spacecraft appear to 
indicate a substantial flux of interstellar particles with masses
$> 10^{-12}\gram$.
The reported abundance of these massive grains cannot be
typical of interstellar gas: it is incompatible with both 
interstellar elemental abundances {\it and}
the observed extinction properties of the interstellar dust population.
We discuss the likelihood that the Solar System is by chance located
near an unusual concentration of massive grains and conclude that 
\newtext{this} 
is unlikely, unless dynamical processes in the ISM are responsible for such
concentrations.  Radiation pressure might conceivably
drive large grains into ``magnetic valleys''.
If the influx direction of interstellar gas and dust is varying
on a $\sim10$~yr timescale, 
\newtext{as suggested by some observations,} 
\newtext{this would}
have dramatic
implications for the small-scale structure of the interstellar medium.

\end{abstract}

\section{\label{sec:intro}
         Introduction}

The interstellar medium (ISM) consists of a partially-ionized, magnetized
gas mixed with
solid particles of dust.  
The ionization state and molecular fraction of
the gas depend primarily 
on the gas density and the local intensity of ultraviolet
radiation that can photodissociate molecules and photoionize molecules and
atoms.
The dust content is determined by the prior history of the gas, including
injection of newly-formed dust in stellar winds and supernova explosions,
grain destruction in violent events such as supernova blast waves,
and grain growth in the interstellar medium 
by both vapor deposition and coagulation in dense regions.

While we do not know the properties of interstellar dust
with precision, they are strongly-constrained by a variety
of observations.  
The observed wavelength dependence of interstellar extinction -- the 
so-called ``reddening curve'' (reviewed in \S~\ref{sec:ism dust}) -- 
provides strong constraints on both
the composition and size distribution of interstellar dust.
In the local regions of the Milky Way, interstellar dust is abundant,
containing a large fraction of the elements (such as Mg, Si, and Fe)
that can be incorporated into refractory solids.
As discussed in \S\ref{sec:abundances}, 
interstellar abundances therefore provide a strong constraint on grain
models.  The size distribution of interstellar grains can be inferred
from the observed average reddening curve together with interstellar
abundance constraints.

Microparticle impacts on detectors on Ulysses and Galileo have
been interpreted as showing a flux of solid particles entering the
Solar system from the local interstellar medium
\citep{Grun+Zook+Baguhl+etal_1993}.
In \S~\ref{sec:abundances} we show that the
population of large
grains inferred from dust impact detectors on Ulysses and Galileo
\citep{Landgraf+Baggaley+Grun+etal_2000,
       Kruger+Landgraf+Altobelli+Grun_2007,
       Kruger+Grun_2008}
is incompatible with average elemental abundances in the ISM.
In \S~\ref{sec:extinction}, we show that such a large grain population would
result in wavelength-dependent extinction very different from 
what is observed.

The Ulysses and Galileo data, if correctly interpreted, imply 
that the Solar System is, by chance, located in a very atypical spot in the
ISM, with an overabundance of very large grains.  
The likelihood of such a scenario is discussed in \S~\ref{sec:atypical}.
In \S~\ref{sec:velocity structure} we comment on 
\newtext{suggestions}
that the interstellar dust inflow vector
\newtext{might have changed} 
appreciably over only $\sim$5~yrs.
Our conclusions
are summarized in \S~\ref{sec:summary}.

\section{\label{sec:ism dust}
         Dust in the Diffuse Interstellar Medium}

In the Milky Way and many other galaxies, a substantial fraction of the
``refractory elements'' in the ISM are in
solid materials, in submicron dust particles.
At least on large scales, 
the dust and gas are well-mixed, with the density of dust
tending to be proportional to the density of gas.

\begin{figure}[h]
  \includegraphics[width=0.75\textwidth,angle=270]{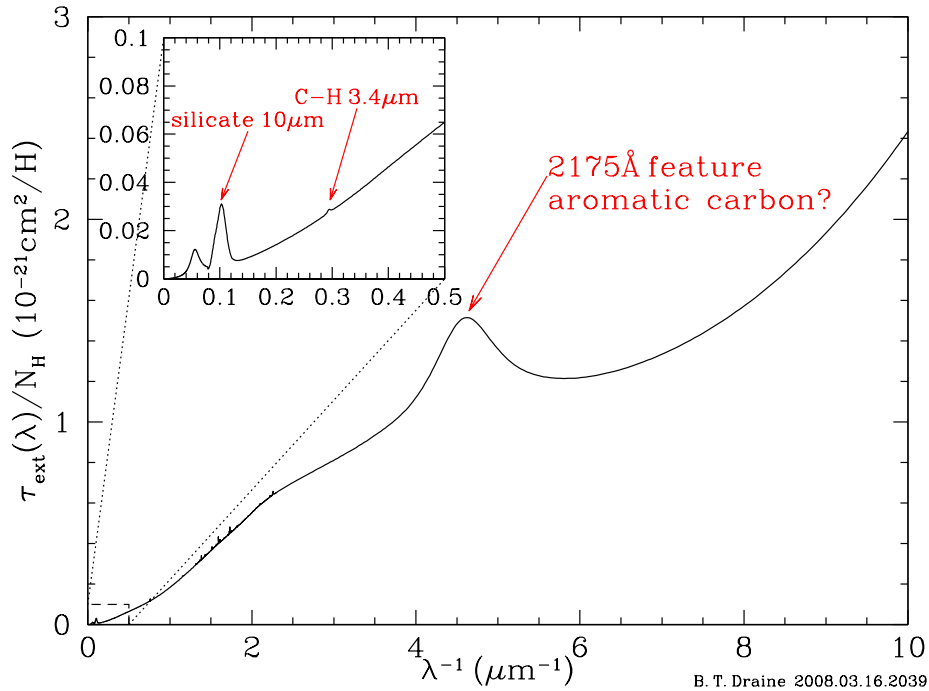}
  \caption{\label{fig:extcurv}
    The average observed extinction per H nucleon, as a function of inverse
    wavelength $1/\lambda$, in diffuse regions of the Milky Way.
    The prominent ``bump'' at $\lambda\approx2175\Angstrom$ is
    probably due to $\pi\rightarrow\pi^*$ electronic transitions
    in $sp^2$-bonded (aromatic) carbon.
    The strong infrared extinction features (see inset) are
    produced by the Si-O stretching mode (9.8\um) and the O-Si-O
    bending mode (18\um).  There is also a weak feature at 3.4\um\ due
    to the C-H stretch in aliphatic (chainlike) hydrocarbons.
    }
\end{figure}

The properties of the dust -- size distribution, shapes, composition --
are inferred from a wide range of observations
\citep[for a review, see][]{Draine_2003a} including: wavelength-dependent
extinction and polarization of starlight, light scattering
in the visible and ultraviolet, small-angle scattering of X-rays,
thermal emission from infrared to
submm wavelengths, and microwave radiation from spinning dust.
Studies of the strength and wavelength-dependence of 
interstellar
extinction 
\newtext{$A_\lambda\equiv (2.5/\ln10)\tau_{\rm ext}(\lambda)
=1.086\tau_{\rm ext}(\lambda)$
}
provide strong constraints on the size
distribution and composition of interstellar dust.
Figure \ref{fig:extcurv} 
shows an empirical parameterization of the extinction by dust in 
``diffuse clouds''
\citep{Cardelli+Clayton+Mathis_1989,
       Fitzpatrick_1999}.
A ``diffuse cloud'' is simply a region with
visual extinction $A_{\rm V}\ltsim 1$~mag;
most of the interstellar H~I is in such regions.
The interstellar material surrounding the heliosphere consists of
diffuse H~I, and it was natural to expect that the interstellar
dust outside the heliosphere would be typical ``diffuse cloud'' dust --
typical in both its size distribution and its abundance relative to the gas.

\newtext{
The wavelength-dependence of $A_\lambda$ is known to vary
from one sightline to another.
Extinction curves are often characterized by
$R_V\equiv A_V/(A_B-A_V)$.
Average diffuse clouds have $R_V\approx 3.1$ but $R_V$ can be
as small as $\sim 2.2$ in some diffuse clouds
\citep[e.g., $R_V=2.22\pm0.14$ toward HD~210121:][]{Fitzpatrick_1999}
and can reach values as large as $\sim5.8$ in dense regions 
\citep[e.g., $R_V=5.8\pm0.6$ toward HD~36982:][]{Fitzpatrick_1999}.
The extinction law shown in Fig.\ \ref{fig:extcurv} is intended to
be an average curve for diffuse clouds, with $R_V\approx 3.1$.}
The most notable characteristic of the extinction
curve in Fig.\ \ref{fig:extcurv}
is the continuing rise into the vacuum ultraviolet; this requires
that the size distribution be such that the
total surface area of the dust is dominated by very small
grains with radii $a\ltsim 200\Angstrom$.
The second notable characteristic is the prominent ``bump'' in the extinction
at $\lambda\approx2175\Angstrom$.  While this feature has not yet been
identified with complete certainty
\citep[see, e.g.,][]{Draine_1989a}, it is thought to be
produced by $\pi\rightarrow\pi^*$  electronic transitions in aromatic
carbon, such as the carbon in graphite or in 
polycyclic aromatic hydrocarbons (PAHs).
The 2175$\Angstrom$ bump 
traces only the aromatic carbon in particles with masses
$m\ltsim 10^{-16}\gram$: the feature is suppressed in larger grains.
The 
\newtext{2175$\Angstrom$}
feature therefore gives only a lower bound on the carbon content of the
dust population: $\gtsim15\%$ of interstellar
carbon is in aromatic structures.

There are also two spectroscopic features in the infrared -- strong absorption
features peaking near $9.7\micron$ and $18\micron$.
These features are characteristic of Si-O stretching
and O-Si-O bending modes in amorphous silicates.  The strength of
the features 
requires that most interstellar Si atoms be incorporated
into these silicates, together with corresponding amounts of Mg, Fe, and O.
Amorphous silicates and carbonaceous materials are 
together thought to
account for the bulk of the mass of interstellar dust in diffuse clouds.
In dense and dark clouds, ices are also present, but the heliosphere is not
located near a dark cloud, hence ices are not expected to be
present in the dust entering
the solar system from the ISM.

As discussed in \S~\ref{sec:abundances}, 
observations of the elements that are ``depleted'' from
the gas phase in interstellar clouds provide an indication of what elements
are in grains -- the bulk of the mass of 
interstellar dust is contributed by
the elements C, O, Mg, Si, and Fe.
Based on the spectroscopic evidence in Figure \ref{fig:extcurv}, it
can be concluded that the dominant materials are some form of amorphous
silicate (with composition $\sim$~MgFeSiO$_4$) and some mixture of carbonaceous
materials -- PAHs, amorphous carbon, graphite, and perhaps even diamond.

\section{\label{sec:abundances}
         Models for Interstellar Dust: Extinction vs. Elemental Abundances}

Observations of the spectra of recently-formed stars, 
together with absorption lines produced
by interstellar gas, have led to estimates of elemental abundances
in the local interstellar material.
Abundances of many elements relative to hydrogen in the ISM
can also be deduced from
emission lines from H~II regions.
Although the Sun was formed out of the ISM 4.5 Gyr ago,
the elemental abundances in the ISM today appear to be close to those
in the solar photosphere, and ``solar abundances'' are generally considered
to be a good guide to interstellar abundances,
although ``solar abundances'' are themselves uncertain: e.g., recent
estimates of O/H in the solar photosphere range from
$(457\pm56)~$ppm \citep{Asplund+Grevesse+Sauval+etal_2004} to
$(730\pm100)~$ppm \citep{Centeno+Socas-Navarro_2008}.
\begin{table}
\begin{center}
\caption{\label{tab:abundances}
         Dust Mass per H from Milky Way Abundances.
$(N_{X}/N_{\rm H})_\odot$ and $(N_{X}/N_{\rm H})_{\rm gas}$ are the abundances
of element X, by number, relative to H in the Sun and in the gas phase of
a ``standard'' interstellar cloud (see text).
$M_{X,{\rm dust}}/M_{\rm H}$ is the mass of element X in dust relative to
the total mass of H.}
\begin{tabular}{l c c c}
\hline\noalign{\smallskip}
$X$ & 
      $(N_{X}/N_{\rm H})_\odot$(ppm)$^a$ &
      $(N_{X}/N_{\rm H})_{\rm gas}/(N_{X}/N_{\rm H})_\odot$~$^a$ &
      $M_{X,{\rm dust}}/M_{\rm H}$ \\
\noalign{\smallskip}\hline\noalign{\smallskip}
C   & 247    & 0.57    & 0.0013 \\
N   & ~~85   & 0.72    & 0.0003 \\
O   & 490    & 0.73    & 0.0021 \\
Mg  & ~~38   & 0.08    & 0.0008 \\
Al  & ~~~~~3$^b$& $\ltsim$0.1$^c$~~~~~~    & 0.0001 \\
Si  & ~~32   & 0.05    & 0.0009 \\
Ca  & ~~~~~2$^b$&~~~~~0.0002$^d$  & 0.0001 \\
Fe  & ~~29   & ~0.007   & 0.0016 \\
Ni  & ~~~~2  & ~0.004   & 0.0001 \\
\noalign{\smallskip}\hline\noalign{\smallskip}
total &   &         & 0.0073 \\
\noalign{\smallskip}\hline\noalign{\smallskip}
\multicolumn{4}{l}{$^a$ \citet{Jenkins_2004} except as noted.}\\
\multicolumn{4}{l}{$^b$ $(N_{X}/N_{\rm H})_\odot$ from \citet{Grevesse+Sauval_1998}}\\
\multicolumn{4}{l}{$^c$ assumed}\\
\multicolumn{4}{l}{$^d$ \citet{Savage+Sembach_1996}}\\
\end{tabular}
\end{center}
\end{table}
The second column of 
Table \ref{tab:abundances} lists the solar abundances of the elements
that are sufficiently abundant 
to contribute 1\% or more of the mass of interstellar
dust.
Elements such as Ti do not appear in Table \ref{tab:abundances}
because they are too
rare: the abundance of Ti, by mass, 
is only about 0.3\% of the abundance of Fe.
Therefore, even though most interstellar
Ti is in fact locked up in grains, Ti is not a major grain constituent.

The third column gives observed gas-phase abundances, relative to solar,
\newtext{in}
``standard'' interstellar diffuse clouds, 
such as the well-studied cloud on the line-of-sight to the bright star
$\zeta$Oph.
The gas-phase abundance of C appears to be only $\sim$57\% of the total
C abundance, implying that $\sim$43\% of the carbon is sequestered
in grains.
For elements such as Mg, Si, or Fe the ``depletions'' are more severe, with
90\% or more of the material locked up in grains.

Based on these observations alone, we can estimate the mass of interstellar
dust: $\sim$0.73\% of the mass of the hydrogen in a ``standard'' cloud.
It is important to recognize that ``solar'' abundances of elements such
as C, Mg, Si, and Fe remain uncertain, and interstellar abundances might
be a bit higher than solar abundances, but it is difficult to imagine
that the total mass of dust in ``standard'' diffuse clouds could be much
more than $\sim$1.0\% of the total hydrogen mass.

\begin{figure}[h]
\begin{center}
  \includegraphics[width=0.75\textwidth,angle=270]{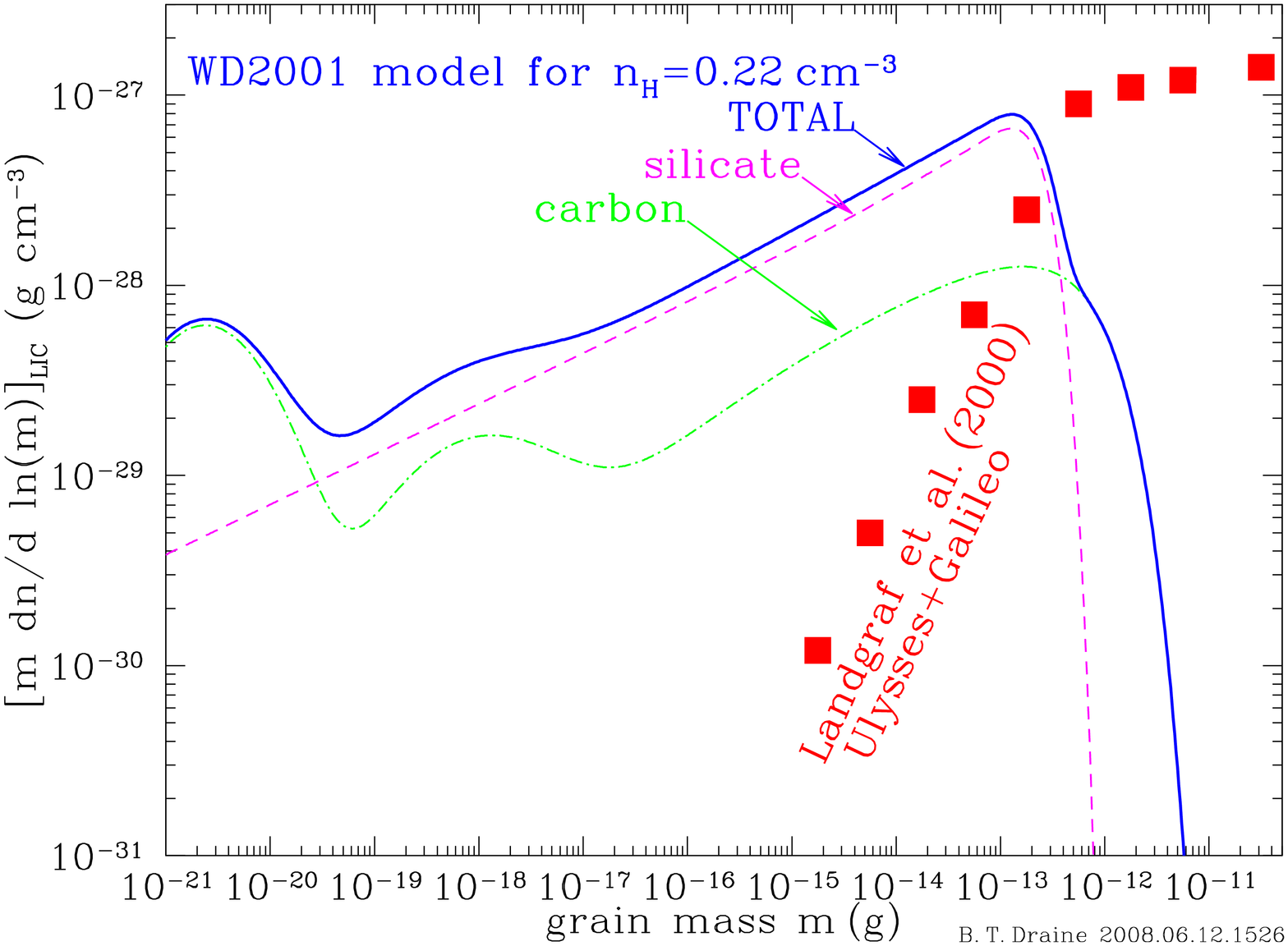}
  \caption{\label{fig:dmda}
         The mass distribution from \citet{Weingartner+Draine_2001a}
         scaled to the density $n_{\rm H}\approx 0.22\cm^{-3}$
	 of the local interstellar cloud.
	 The peak near $\sim3\times10^{-21}\gram$ consists of PAHs.
         Also shown is the mass distribution estimated from
	 impacts on Ulysses and Galileo
	 \citep{Landgraf+Baggaley+Grun+etal_2000}.
	 No correction for ``filtration'' by the heliospheric magnetic field
	 has been applied.
	 For 
	 $5\times 10^{-13} < m < 3\times10^{-11}\gram$ 
	 the mass flux observed by Ulysses and Galileo is far above
	 that expected for interstellar dust (see text).}
\end{center}
\end{figure}

Various authors have obtained dust grain size distributions that
reproduce the observed extinction per H as shown in Figure \ref{fig:extcurv}, 
subject to the constraint that the
mass of the dust in the model should be consistent with the ``observed''
mass given in Table \ref{tab:abundances}
\citep[e.g.,][]{Mathis+Rumpl+Nordsieck_1977,
                Draine+Lee_1984,
                Weingartner+Draine_2001a,
		Zubko+Dwek+Arendt_2004}.
This turns out not to be an easy task: models that reproduce the observed
extinction -- even when trying to also minimize the total grain mass -- tend
to consume 100\% or more of the ``available'' material.
Modest discrepancies between the mass in the dust model and the
``observed'' dust mass in Table \ref{tab:abundances} would not be
unexpected, given uncertainties in the observations, and given that the
theoretical models make simplifying assumptions, e.g., typically assuming
spherical grains.
Overall, one draws the conclusion that the bulk of the interstellar grain
mass is in dust grains with masses $\ltsim5\times10^{-13}\gram$ -- these
grains are {\it needed} to produce the observed extinction, and there isn't
much dust mass ``left over'' once the observed extinction has been reproduced.

Assuming that the interstellar grain population consists of two distinct
compositions -- amorphous silicate grains and carbonaceous grains --
\citet[][hereafter WD01]{Weingartner+Draine_2001a} found size distributions
for these two components
that would produce extinction close to the observed extinction curve
in Fig.\ \ref{fig:extcurv}, and which would incorporate amounts of C, Mg, Si,
and Fe approximately consistent with current estimates of elemental
abundances in the ISM.
The same dust model, heated by starlight, is consistent with observations
of infrared emission from the Milky Way and similar galaxies
\citep{Draine+Li_2007,Draine+Dale+Bendo+etal_2007}.
The resulting mass distributions are shown in Fig.\ref{fig:dmda},
for an H nucleon density $n_{\rm H}=0.22\cm^{-3}$, 
the value currently estimated
for the very local ISM based on observations of inflowing He$^0$
\citep{Lallement+Raymond+Vallerga+etal_2004}
and photoionization models for the nearby ISM
\citep{Slavin+Frisch_2007}.

\begin{figure}[h]
\begin{center}
  \includegraphics[width=0.75\textwidth,angle=270]{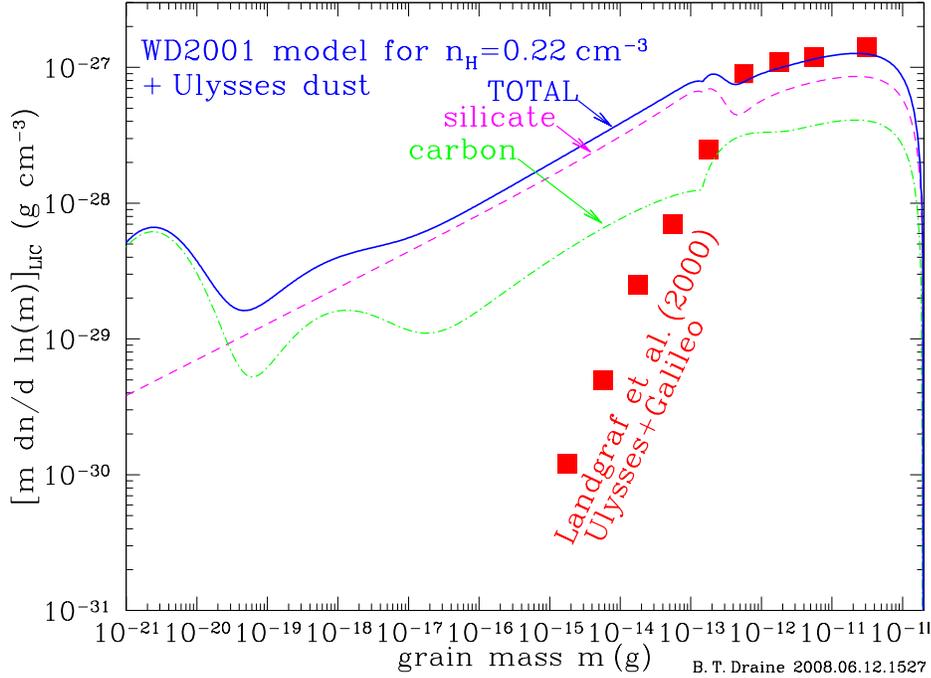}
  \caption{\label{fig:dmda_big}
         The mass distribution from \citet{Weingartner+Draine_2001a}
	 plus the ``big grain'' component from the Ulysses and Galileo
	 measurements.  This model has a dust/H mass ratio of 0.028,
	 much larger than the value 0.010 of the WD01 size distribution
         in Fig.\ \ref{fig:dmda}.}
\end{center}
\end{figure}
\begin{figure}[h]
\begin{center}
  \includegraphics[width=0.75\textwidth,angle=270]{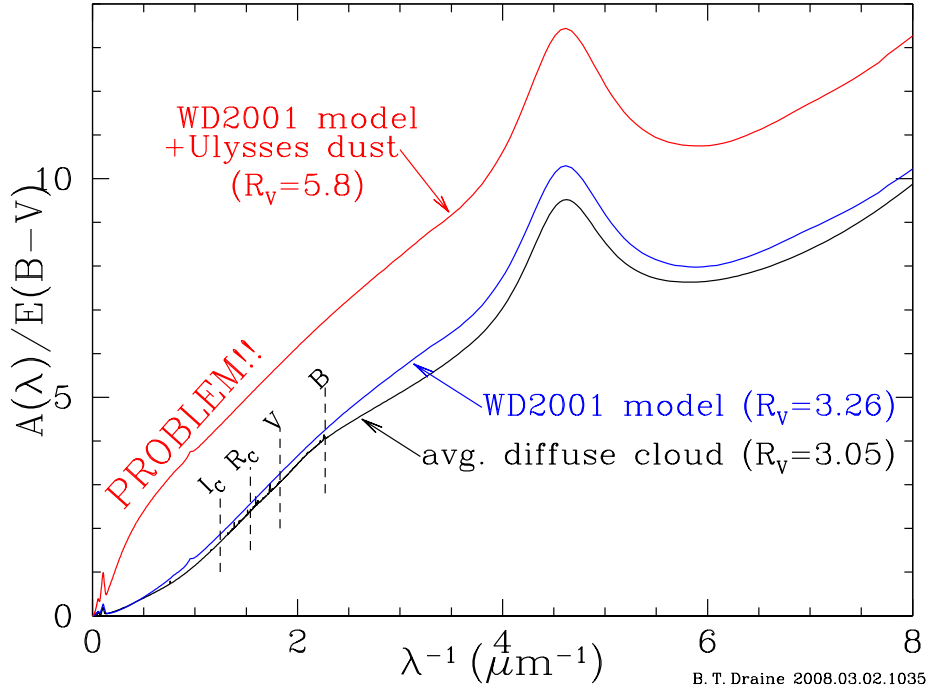}
  \caption{\label{fig:extcurv_big}
         Reddening law calculated for the mass distribution of 
	 Fig.\ \ref{fig:dmda_big} 
	 (note: $E(B-V)\equiv A(B)-A(V)$,
	 where $A(\lambda)$ is the extinction at wavelength $\lambda$).
	 The large grains contribute substantial amounts of additional
	 extinction, and
	 the resulting reddening curve differs strongly from observed
	 reddening: 
	 $A(\lambda)/E(B-V)$ exceeds observed values by 
	 factors $1.5-2$ for the commonly-observed B, V, R, and I bands.
         }
\end{center}
\end{figure}

Also shown in Fig.\ \ref{fig:dmda} is the mass distribution of
particles entering the heliosphere from the local ISM
as estimated by \citet{Landgraf+Baggaley+Grun+etal_2000} from the dust
impact detectors on Ulysses and Galileo.
Because the magnetic field of the heliosphere is expected to substantially
deflect incoming particles with masses $m\ltsim 3\times10^{-13}\gram$, the
fact that the \citet{Landgraf+Baggaley+Grun+etal_2000} results
fall well below the WD2001 model for $m<10^{-13}\gram$ is not surprising.
However, the reported flux of $m \gtsim 3\times10^{-13}\gram$ particles
is quite unexpected if the local ISM has a dust/gas ratio typical
of diffuse regions in our Galaxy.

First of all, there is the question of overall mass: as seen from
Table \ref{tab:abundances}, current estimates for solar and interstellar
abundances would allow $M_{\rm dust}/M_{\rm H}$ of only 0.0073.
Given uncertainties in both measured abundances and grain modeling, 
it can be argued that the WD2001 dust
model ($M_{\rm dust}/M_{\rm H} \approx 0.010$ -- a factor 1.4 greater
than the total in Table \ref{tab:abundances}) is within tolerances.
However, extending the size distribution to include the Ulysses results,
as in Fig.\ \ref{fig:dmda_big}, raises $M_{\rm dust}/M_{\rm H}$
to 0.028 -- 
3.9 times higher than the estimated total in Table \ref{tab:abundances}.
This is incompatible with our current understanding 
of elemental abundances in the general ISM.

\section{\label{sec:extinction}
         Contribution of Massive Grains to Extinction}

If the massive grains detected by \citet{Landgraf+Baggaley+Grun+etal_2000}
were part of the general interstellar grain population, they would have
conspicuous effects on the interstellar extinction.
To see this, we have taken the WD01 size distribution, and added to it
an additional population of carbonaceous and silicate particles so
as to approximately reproduce the \citet{Landgraf+Baggaley+Grun+etal_2000}
size distribution at $m>3\times10^{-13}\gram$.
We arbitrarily assume that
2/3 of the added mass
is contributed by amorphous silicates and 1/3 by graphite.
The adopted size distribution is shown in Fig.\ \ref{fig:dmda_big}.
Approximating the particles as spheres, the extinction as a function
of wavelength has been calculated for the extended size distribution of
Figure \ref{fig:dmda_big}.
The resulting ``reddening curve'' $A(\lambda)/E(B-V)$
is shown in Fig.\ \ref{fig:extcurv_big}.

On suitable sightlines, $A(\lambda)/E(B-V)$ can be determined 
observationally to accuracies
of $\sim10\%$ for $0.5\ltsim (\lambda/\micron)^{-1}\ltsim 3$.
The reddening law shown in Fig.\ \ref{fig:extcurv_big}
is well outside the range of what is observed 
\citep[see, e.g.][]{Mathis_1990}.
\newtext{The synthetic curve in Fig.\ \ref{fig:extcurv_big} has 
$R_V\approx5.8$ -- such large values of $R_V$ are not seen in diffuse clouds,
being found only in regions with $A_V\gtsim 2$.}

{\it It does not seem possible for the dust in the general ISM to have the
size distribution for $m \gtsim 3\times10^{-13}\gram$ 
reported by \citet{Landgraf+Baggaley+Grun+etal_2000}:}
(1) as shown in \S\ref{sec:abundances}, 
there are simply not enough atoms of C, Mg, Si, and Fe to constitute
such a large mass in dust, and,
(2) as seen here, 
if such dust were pervasive, the wavelength-dependence of interstellar
extinction would be totally unlike what is actually observed.

\section{\label{sec:atypical}
         Could the Dust in the Local ISM Be Atypical?}
We have shown above that the large-grain population reported by
\citet{Landgraf+Baggaley+Grun+etal_2000} cannot be pervasive.
However, 
it is important to realize that the dust detectors on Ulysses and
Galileo have only probed a tiny portion of the ISM: a cylindrical volume
with diameter $\sim$10~AU, and length increasing by $\sim$5~AU/yr
due to the solar-system's motion of 26.2~km/s relative to the local ISM
\citep{Mobius+Bzowski+Chalov+etal_2004}.  
We have therefore probed only a ``microscopic''
sample of the ISM -- how representative do we expect this sample to be?

\newtext{\subsection{Turbulent Mixing in the ISM}}

MHD turbulence appears to be pervasive in the ISM.  Although not understood
in detail, the turbulence appears to be the result of ``driving'' by
energetic phenomena on large scales $L_{\rm max}$-- e.g., 
H~II regions, stellar winds, supernova explosions.
The turbulent cascade to smaller scales appears to approximately follow
the ``Kolmogorov'' power-law scaling, with the velocity differences on
length-scale $L$ varying as
\beq \label{eq:kolmogorov}
v_L \approx v_{\rm max}(L/L_{\rm max})^{1/3}~~~~~{\rm for~~~}
L_{\rm diss} < L < L_{\rm max} ~~~,
\eeq
where $L_{\rm diss}$ is the
length scale below which dissipation is dominant.
Because of the magnetic field, the turbulence is anisotropic, but the
scaling law (\ref{eq:kolmogorov}) approximately applies to turbulent
motions perpendicular to the magnetic field.

Observations of turbulence within $\sim100\pc$ are more-or-less
consistent with $v_{\rm max}\approx 10\kms$ and $L_{\rm max}\approx 100\pc$.
Nonuniformities on a scale $L$ will be erased on timescales
\beq
\tau_{\rm diff}\approx \frac{L}{v_L} 
\approx \frac{L^{2/3}L_{\rm max}^{1/3}}{v_{\rm max}}
\approx 0.5~\left(\frac{L}{\pc}\right)^{2/3} \Myr ~~~,
\eeq
where we have adopted $v_{\rm max}\approx 10\kms$ and 
$L_{\rm max}\approx 100\pc$.
A mixing timescale of $\ltsim$Myr is short relative to
Galactic timescales.
Therefore we do not expect to find small-scale abundance inhomogeneities
unless they were very recently injected, or unless some specific mechanism
sustains them.
What injection mechanisms might produce local enhancements in the
population of large dust particles?

\subsection{Enrichment by Supernova Explosions?}

One possible source of inhomogeneity is Type II supernova
explosions following core collapse in massive stars.  Each such explosion 
enriches the nearby ISM with $\sim 5\Msol$ of heavy elements, a fraction of
which may be in grains.  Hydrodynamic instabilities in the supernova remnant
will mix these heavy elements with a mass $M_{\rm mix}$ of the ISM.
The normal ISM has a heavy-element mass fraction $Z\approx 0.02$;
this will be enhanced by $\Delta Z \approx 0.05(10^2\Msol/M_{\rm mix})$.
For the average density $\langle n_{\rm H}\rangle\approx 1\cm^{-3}$ of
the ISM in the solar neighborhood, this corresponds to a lengthscale
$L_{\rm mix} \approx 14 \pc (M_{\rm mix}/10^2\Msol)^{1/3}$ 
and from eq. (2), we would
expect inhomogeneities on this length scale to be erased in a time
\beq \label{eq:tau_diff}
\tau_{\rm diff} \approx 2.9 (M_{\rm mix}/10^2\Msol)^{2/9} \Myr
\eeq
The supernova rate/volume in the Galactic disk 
is $S\approx 10^{-13}\pc^{-3}\yr^{-1}$.
The probability that a SN exploded within a distance $L_{\rm mix}$ within
a time $\tau_{\rm diff}$ is only 
$\sim L_{\rm mix}^3 S \tau_{\rm diff}
\approx 10^{-3}$.
It is therefore very improbable that the local interstellar
cloud has been heavily enriched by a recent SN explosion.  The Local Bubble
is believed to have been caused by one or more SN explosions over
the past 10--15~Myr, but
these were located at a distance of $\sim100\pc$
\citep{Fuchs+Breitschwertd+deAvillez+etal_2006}.
The strongest argument against enrichment by 
\newtext{SN ejecta is}
the fact that the gas-phase abundances of Mg and Fe appear to show
normal depletions relative to solar abundances
\citep{Redfield+Linsky_2008}.

\subsection{Wake of an Evolved Star?}
Cool AGB stars have dusty winds that may pollute the ISM with fresh
grain material -- for example, the wind from Mira~=~$\omicron$~Ceti
\citep{Martin+Seibert+Neill+etal_2007}.  What is the likelihood that
a recent passage by an AGB star left behind a concentration of
large grains that might account for the excess of large particles
seen by Ulysses?

Consider a star moving at speed $v_\star$ relative to the ISM, losing mass
at a rate $\dot{M}$.  It will leave behind a wake, with radius $R_w$,
filled with gas with density $n_w$ and temperature $T_w$.
Mass conservation 
and balance with the interstellar pressure $p_{\rm ISM}$ give
\beq
n_w 1.4 m_{\rm H} \pi R_w^2 v_\star = \dot{M} ~~~,
\eeq
\beq
n_w k T_w = p_{\rm ISM} ~~~~~.
\eeq
These two equations can be solved for the wake radius
\beqa
R_w = 
\left[
\frac{\dot{M}}{1.4m_{\rm H}\pi v_\star}
\frac{T_w}{p_{ISM}/k}
\right]^{1/2}
&=& 0.13 \pc
\left[
\frac{\dot{M}_{-6}}{v_{\star,6}}
\frac{T_{w2}}{(p_{\rm ISM}/k)_{5000}} 
\right]^{1/2} ~~~,
\\ \nonumber
\dot{M}_{-6}\equiv \frac{\dot{M}}{10^{-6}\Msol\yr^{-1}}
~~~~~~&,&~~~~~~
v_{\star,6} \equiv \frac{v_\star}{10\kms} ~~~,
\\ \nonumber
T_{w2}\equiv \frac{T}{100\K}
~~~~~~&,&~~~~~
\left(\frac{p_{\rm ISM}}{k}\right)_{5000} 
\equiv \frac{p_{\rm ISM}/k}{5000\cm^{-3}\K} ~~~.
\eeqa
The trailing wake will be mixed with the ISM by turbulent diffusion on 
a time given by eq.\ (\ref{eq:tau_diff}):
\beq
\tau_{\rm diff}
\approx 
0.5 (R_w/\pc)^{2/3}\Myr
\approx 0.13 
\left[
\frac{\dot{M}_{-6}}{v_{\star,6}}
\frac{T_{w2}}{(p_{\rm ISM}/k)_{5000}}
\right]^{1/3}\Myr ~~~.
\eeq
If the duration of the mass loss phase is longer than $\tau_{\rm diff}$,
the wake volume will be
\beq
V_w 
\approx 
\pi R_w^2 v_\star \tau_{\rm diff}
\approx 0.07 
\left[\frac{\dot{M}_{-6} T_{w,2}}{(p_{\rm ISM}/k)_{5000}}\right]^{4/3}
 v_{\star,6}^{-1/3}\pc^3 ~~~.
\eeq
The {\it total} rate of stellar mass loss in the Milky Way is
$\sim1\Msol\yr^{-1}$, e.g., $10^6$ stars, each with $\dot{M}_{-6}\approx 1$.
If $\sim10^6$ 
stars are randomly-distributed in a disk of full-thickness $\sim200\pc$
and radius $12\kpc$, then the stellar density is 
$n_\star\approx 1\times10^{-5}\pc^{-3}$, and the nearest-neighbor distance is
$n_\star^{-1/3}\approx 50\pc$.  
[The distance to Mira, $D=107\pc$, is in rough agreement with our estimate
for $n_\star^{-1/3}$.]
With $n_\star\approx10^{-5}\pc^{-3}$,
the fraction of the volume occupied by ``wakes'' is very small:
\beq
n_\star V_w \approx 7\times10^{-7}
\left[\frac{\dot{M}_{-6} T_{w,2}}{(p_{\rm ISM}/k)_{5000}}\right]^{4/3}
 v_{\star,6}^{-1/3} ~~~.
\eeq
It is therefore {\it extremely} unlikely that the Solar System
would by chance be located today within such a stellar wake; this conclusion
will not be changed for any plausible variation of uncertain parameters
such as $\dot{M}_{-6}$, $T_{w,2}$, or $v_{\star,6}$

\subsection{\label{subsec:dynamical}
            Dynamical Concentration of Massive Grains?}

We have seen above that it is highly unlikely that the Solar
System is by chance passing through gas that was recently
enriched in very large grains from either a supernova explosion or
an evolved star.  What other process might produce the anomalous
concentration of dust grains that appears to be present in the
portion of the ISM we are now passing through?

\begin{figure}[h]
\begin{center}
  \includegraphics[width=0.45\textwidth,angle=0]{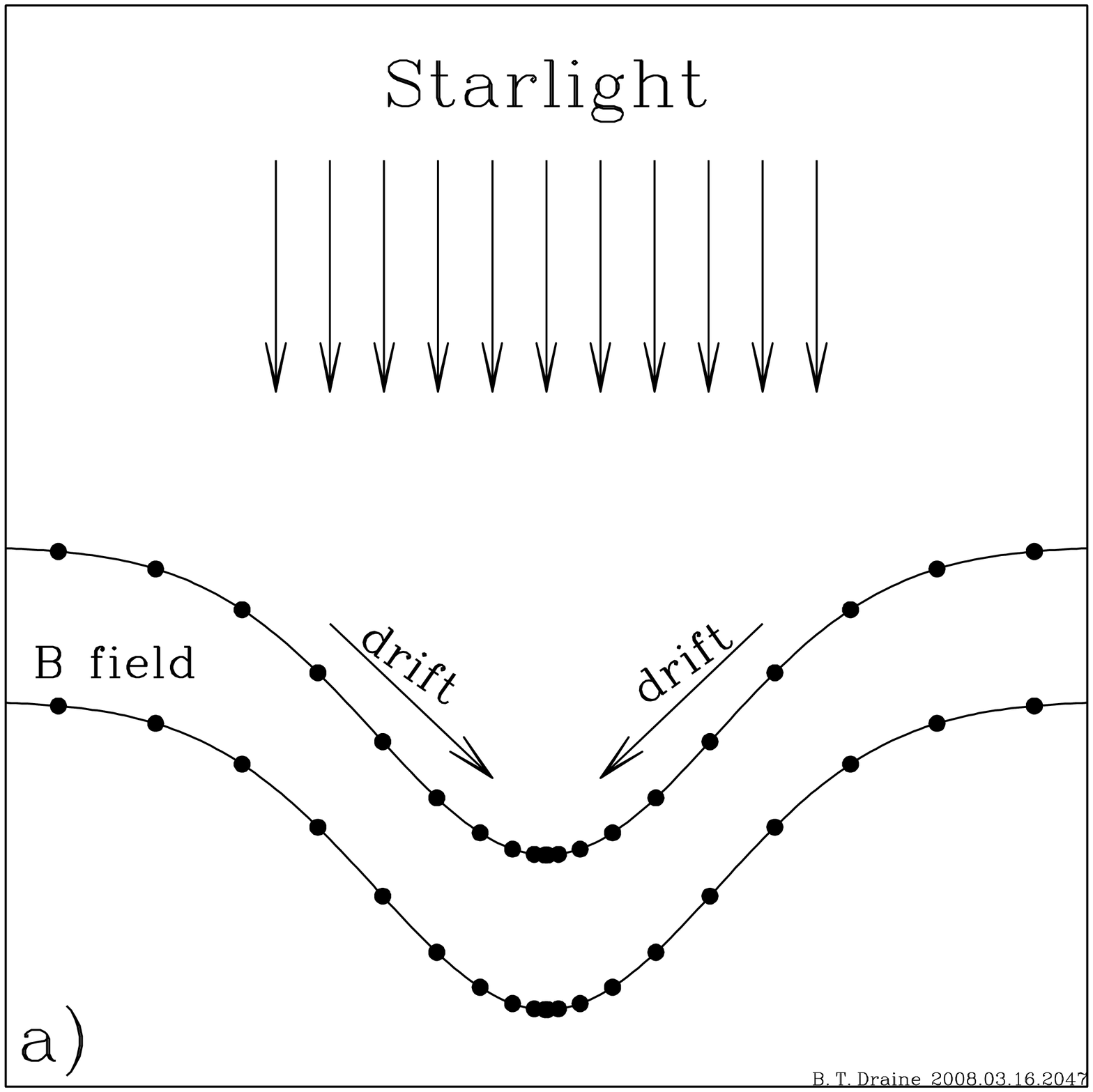}
  \includegraphics[width=0.45\textwidth,angle=0]{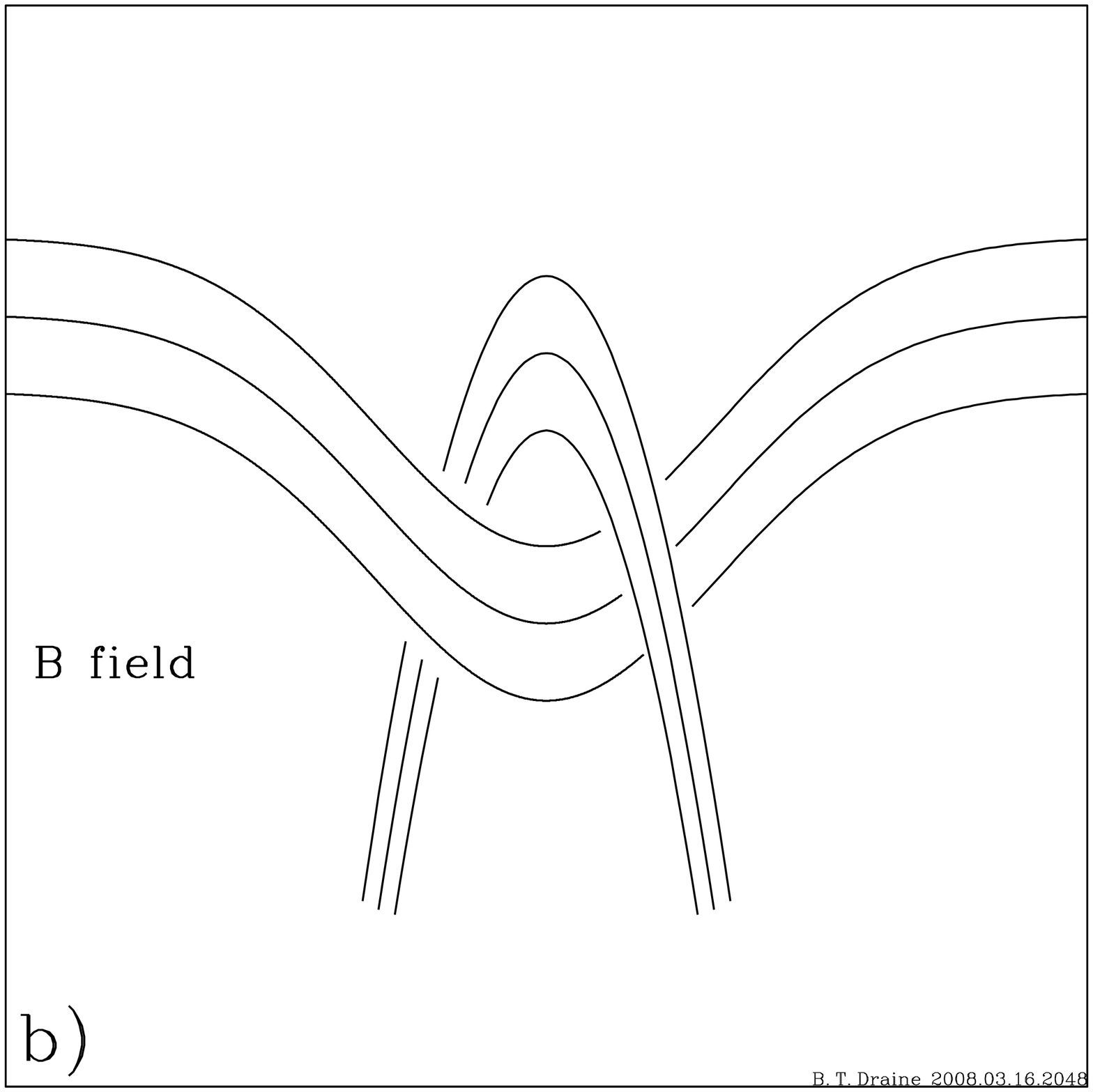}
  \caption{\label{fig:drift}
           (a) Radiation pressure-driven drift of dust grains along
	   magnetic field lines could concentrate grains in
	   magnetic ``valleys'' (see text).
	   (b) Tangled magnetic fields could produce magnetic valleys.
	   }
\end{center}
\end{figure}
Dust grains and gas atoms are
subject to different forces, and in general the dust grains will 
{\it drift} relative to the gas.  Charged dust grains are coupled to the
magnetic field, which inhibits drift across magnetic field lines, but
the grains are free to drift {\it along} field lines.
Drift velocities resulting from radiation pressure and other effects
of anisotropic starlight have been discussed by
\citet{Weingartner+Draine_2001b}.  The drift velocities are not large, but
can attain $\sim 0.5\kms$ in the ``warm neutral medium'' conditions
characteristic of the region the Solar System is now moving through
\citep[see Figs.\ 17,18 of][]{Weingartner+Draine_2001b}.
If sustained for long enough, these drifts 
might result in variations in the dust/gas ratio.

One possible scenario for concentrating dust
is illustrated in Fig.\ \ref{fig:drift}a.
If field lines are bent,
radiation pressure could push grains
into magnetic ``valleys'', as shown.
If the width and depth of the ``valley'' are
both of size $L_B$, then
dust might accumulate on a time scale
\beq
t_{\rm accum}\approx
\frac{L_B}{v_{\rm drift}}
\approx 10^3\yr
\left(\frac{L_B}{100\AU}\right)
\left(\frac{0.5\kms}{v_{\rm drift}}\right) ~~~.
\eeq
Magnetic stresses will act to try to straighten the field
lines.
Radiation pressure acting on the grains, if strong enough, could
keep the field deformed, and could even cause the field deformation
to grow, in a manner
akin to the Parker instability, except with radiation pressure on dust
playing the role of gravity on gas.
However, this would require balancing the magnetic force per volume
$\sim\nabla(B^2/8\pi)\approx B^2/8\pi L_B$ with the radiation pressure
force per volume $\kappa \rho J_{\rm rad}/c$, where
$\kappa$ is the dust opacity, $\rho$ is the dust mass density,
and $J_{\rm rad}$ is the net flux of starlight.
With the magnetic force/volume scaling as $1/L_B$, and parameters appropriate
to the Milky Way,
it does not appear that radiation pressure on dust could deform the
magnetic field on length scales $L_B\ltsim 10\pc$.
However, local field curvature might be maintained by magnetic stresses
if the magnetic field is tangled, as shown in Fig.\ \ref{fig:drift}b.

If radiation-pressure-driven drift is 
responsible for concentrating very large grains,
it should also have acted to concentrate the $m\approx 10^{-13}\gram$
grains that are thought to dominate the grain size distribution in the
average interstellar medium
(recall Fig.\ \ref{fig:dmda}), as their drift velocities will be similar
to those of larger grains.  
It is not clear that the ``filtration''
effects in the heliosphere will be able to suppress the flux of these
particles to the values observed by Ulysses and Galileo.

\section{\label{sec:velocity structure}
         Structure of the Very Local ISM}

Recent analyses of microparticle impacts on the Ulysses spacecraft appear
to indicate that the 
\newtext{impacting}
dust velocity vector in heliocentric
coordinates has shifted by 30$^\circ$ over the 15 years of observation
\citep{Kruger+Landgraf+Altobelli+Grun_2007}.
The interstellar dust mass flux 
\newtext{at 4-5~AU} 
also appears to have varied by a factor $\sim$3 over 1992-2006. 
\newtext{These variations might be due to solar-cycle-related changes
in the interplanetary B field at $\gtsim5\AU$
\citep{Landgraf+Kruger+Altobelli+Grun_2003,
       Kruger+Landgraf+Altobelli+Grun_2007},
but variations in such electromagnetic ``filtration'' would be
expected to result in variations in velocity vector and flux as a function
of grain size, with electromagnetic deflection expected to be minimal for
$m \gtsim 2\times10^{-12}\gram$.
Surprisingly, size-dependence of the velocity vector is not evident in the
data \citep{Kruger+Landgraf+Altobelli+Grun_2007}, so we must consider the
possibility that the grain mass flux impinging on the heliosphere is
variable.}
Since 15 years of observation corresponds to a spatial scale of only
$\sim$83~AU, 
\newtext{variations in the grain flux incident on the heliosphere would
require}
substantial
variations in both grain density and velocity over length scales of only
tens of AU.
Such small-scale variations in the dust density
in the local interstellar medium, if present,
would appear to require an active mechanism, such as described in
\S\ref{subsec:dynamical}, to maintain it.  While \newtext{slow}
dust drift 
\newtext{relative to the gas}
might account for
density variations, one would not expect large velocity variations in
a quiescent medium 
\citep[][estimated $v_{\rm drift}\ltsim 0.5\kms$]{Weingartner+Draine_2001b}.

It is interesting to note that the velocity of the inflowing He$^0$ does not
coincide with the velocity of two closest interstellar clouds:
the ``Local Interstellar Cloud'' (LIC) and ``Cloud G'':
the velocity of the local He$^0$ is close to the {\it average} of the LIC and G
cloud velocity vectors
\citep{Redfield+Linsky_2008}.
In view of this, it is natural to consider the possibility 
that the heliosphere might,
by chance, be located in the narrow shock transition where the two clouds
interact: the time-dependence of the mass flux and velocity 
of inflowing atoms and dust grains may be revealing structure in
a multifluid shock transition layer.

\section{\label{sec:summary}
         Summary}

The size distribution of interstellar grains entering the heliosphere,
as inferred from observations by Ulysses and Galileo 
\citep{Landgraf+Baggaley+Grun+etal_2000,Kruger+Landgraf+Altobelli+Grun_2007}
cannot be typical of the general interstellar medium, as can be
demonstrated by two independent arguments:
\begin{enumerate}
\item The required
abundance of elements in grains would substantially exceed what is
available in the interstellar medium.
\item If such a size distribution were generally present, it would
produce an interstellar ``reddening law'' very different from what is
observed.
\end{enumerate}
Therefore, if the size distribution of local interstellar dust does
have the large grain population reported by 
\citet{Landgraf+Baggaley+Grun+etal_2000}, 
the dust grain/gas ratio in the interstellar medium must be quite
nonuniform.  The length scale characterizing these nonuniformities
is not known.  If the velocity vector of the incoming dust flow is
\newtext{actually}
changing over time scales of only years
\newtext{ -- one possible explanation for the variations in the directions
of impacting particles reported by}
\citet{Kruger+Landgraf+Altobelli+Grun_2007} 
\newtext{--} this would require
that the dust velocity vary over lengthscales of only tens of AU.
Such small scale structure was not expected.

Mechanisms that might account for such nonuniformity are considered.
It seems extremely unlikely that the Sun is passing through a region that
has recently been enriched with dust from a stellar source.
The least unlikely scenario may involve concentration of dust
in certain regions, and removal of dust from other regions, by dynamical
processes.  One possible mechanism involving anisotropic starlight
driving dust grains along deformed magnetic field lines is outlined.
Whether this can compete with the diffusive effects of turbulent
mixing is far from clear, however.

It is important to carry out additional observations to confirm the
enhanced grain size distribution, and to confirm the time-dependence of
the density and velocity vector of the inflowing dust and gas.
If the reported density of large grains, and the time-dependence of
the inflow, are confirmed, this may require revision of our
understanding of the small-scale structure of the ISM.
Absorption line studies seem to suggest that, 
by coincidence, the heliosphere is
just now passing through the transition zone -- possibly a shock transition --
between the ``Local Interstellar Cloud'' and ``Cloud G''.  If so, the flow
into the heliosphere offers the opportunity to study the small-scale
structure in this transition zone.
The Ulysses observations indicate that
this region is heavily enriched with large dust particles,
although why this should be so remains unclear.


\begin{acknowledgements}
I thank the organizers for inviting me to participate in the Workshop.
I am grateful to R.H. Lupton for availability of the SM graphics program,
used in preparation of the figures in this paper.  I thank
E. Gr\"un, H.~Kr\"uger, J.~Linsky, and an anonymous referee 
for helpful discussions and comments.
This research was supported in part by NSF grant AST-0406883.
\end{acknowledgements}

\newcommand{\aap}{{A\&A\ }}
\newcommand{\apj}{{ApJ\ }}
\newcommand{\apjl}{{ApJ\ }}
\newcommand{\apjs}{{ApJ~Suppl.\ }}
\newcommand{\araa}{{ARAA\ }}
\newcommand{\jgr}{{JGR\ }}
\newcommand{\mnras}{MNRAS\ }
\newcommand{\nature}{{Nature\ }}
\newcommand{\pasp}{{PASP\ }}
\newcommand{\bibfont}{\footnotesize}
\bibliographystyle{SSRv}

\begin{thebibliography}{28}
\expandafter\ifx\csname natexlab\endcsname\relax\def\natexlab#1{#1}\fi

\bibitem[{{Asplund} et~al.(2004){Asplund}, {Grevesse}, {Sauval}, {Allende
  Prieto}, and {Kiselman}}]{Asplund+Grevesse+Sauval+etal_2004}
M.~{Asplund}, N.~{Grevesse}, A.~J. {Sauval}, C.~{Allende Prieto},
  D.~{Kiselman}, \aap {\bf 417}, 751--768 (2004)

\bibitem[{{Cardelli} et~al.(1989){Cardelli}, {Clayton}, and
  {Mathis}}]{Cardelli+Clayton+Mathis_1989}
J.~A. {Cardelli}, G.~C. {Clayton}, J.~S. {Mathis}, \apj {\bf 345}, 245--256
  (1989)

\bibitem[{{Centeno} and {Socas-Navarro}(2008)}]{Centeno+Socas-Navarro_2008}
R.~{Centeno}, H.~{Socas-Navarro}, \apjl {\bf 682}, L61--L64 (2008)

\bibitem[{{Draine}(1989)}]{Draine_1989a}
B.~T. {Draine}, In {\em IAU Symp. 135: Interstellar Dust\/}, edited by
  L.~Allamandola, A.~Tielens, pages 313--327 (1989)

\bibitem[{{Draine}(2003)}]{Draine_2003a}
B.~T. {Draine}, \araa {\bf 41}, 241--289 (2003)

\bibitem[{{Draine} et~al.(2007){Draine}, {Dale}, {Bendo}, {Gordon}, {Smith},
  {Armus}, {Engelbracht}, {Helou}, {Kennicutt}, {Li}, {Roussel}, {Walter},
  {Calzetti}, {Moustakas}, {Murphy}, {Rieke}, {Bot}, {Hollenbach}, {Sheth}, and
  {Teplitz}}]{Draine+Dale+Bendo+etal_2007}
B.~T. {Draine}, D.~A. {Dale}, G.~{Bendo}, K.~D. {Gordon}, J.~D.~T. {Smith},
  L.~{Armus}, C.~W. {Engelbracht}, G.~{Helou}, R.~C. {Kennicutt}, A.~{Li},
  H.~{Roussel}, F.~{Walter}, D.~{Calzetti}, J.~{Moustakas}, E.~J. {Murphy},
  G.~H. {Rieke}, C.~{Bot}, D.~J. {Hollenbach}, K.~{Sheth}, H.~I. {Teplitz},
  \apj {\bf 663}, 866--894 (2007)

\bibitem[{{Draine} and {Lee}(1984)}]{Draine+Lee_1984}
B.~T. {Draine}, H.~M. {Lee}, \apj {\bf 285}, 89--108 (1984)

\bibitem[{{Draine} and {Li}(2007)}]{Draine+Li_2007}
B.~T. {Draine}, A.~{Li}, \apj {\bf 657}, 810--837 (2007)

\bibitem[{{Fitzpatrick}(1999)}]{Fitzpatrick_1999}
E.~L. {Fitzpatrick}, \pasp {\bf 111}, 63--75 (1999)

\bibitem[{{Fuchs} et~al.(2006){Fuchs}, {Breitschwerdt}, {de Avillez},
  {Dettbarn}, and {Flynn}}]{Fuchs+Breitschwertd+deAvillez+etal_2006}
B.~{Fuchs}, D.~{Breitschwerdt}, M.~A. {de Avillez}, C.~{Dettbarn}, C.~{Flynn},
  \mnras {\bf 373}, 993--1003 (2006)

\bibitem[{{Grevesse} and {Sauval}(1998)}]{Grevesse+Sauval_1998}
N.~{Grevesse}, A.~J. {Sauval}, Space Science Reviews {\bf 85}, 161--174 (1998)

\bibitem[{{Grun} et~al.(1993){Grun}, {Zook}, {Baguhl}, {Balogh}, {Bame},
  {Fechtig}, {Forsyth}, {Hanner}, {Horanyi}, {Kissel}, {Lindblad}, {Linkert},
  {Linkert}, {Mann}, {McDonnell}, {Morfill}, {Phillips}, {Polanskey},
  {Schwehm}, {Siddique}, {Staubach}, {Svestka}, and
  {Taylor}}]{Grun+Zook+Baguhl+etal_1993}
E.~{Grun}, H.~A. {Zook}, M.~{Baguhl}, A.~{Balogh}, S.~J. {Bame}, H.~{Fechtig},
  R.~{Forsyth}, M.~S. {Hanner}, M.~{Horanyi}, J.~{Kissel}, B.-A. {Lindblad},
  D.~{Linkert}, G.~{Linkert}, I.~{Mann}, J.~A.~M. {McDonnell}, G.~E. {Morfill},
  J.~L. {Phillips}, C.~{Polanskey}, G.~{Schwehm}, N.~{Siddique}, P.~{Staubach},
  J.~{Svestka}, A.~{Taylor}, \nature {\bf 362}, 428--430 (1993)

\bibitem[{{Jenkins}(2004)}]{Jenkins_2004}
E.~B. {Jenkins}, In {\em Origin and Evolution of the Elements\/}, edited by
  A.~{McWilliam}, M.~{Rauch}, pages 336--353 (2004)

\bibitem[{{Krueger} and {Gruen}(2008)}]{Kruger+Grun_2008}
H.~{Krueger}, E.~{Gruen}, ArXiv e-prints {\bf 802} (2008)

\bibitem[{{Kr{\"u}ger} et~al.(2007){Kr{\"u}ger}, {Landgraf}, {Altobelli}, and
  {Gr{\"u}n}}]{Kruger+Landgraf+Altobelli+Grun_2007}
H.~{Kr{\"u}ger}, M.~{Landgraf}, N.~{Altobelli}, E.~{Gr{\"u}n}, Space Science
  Reviews {\bf 130}, 401--408 (2007)

\bibitem[{{Lallement} et~al.(2004){Lallement}, {Raymond}, {Vallerga},
  {Lemoine}, {Dalaudier}, and {Bertaux}}]{Lallement+Raymond+Vallerga+etal_2004}
R.~{Lallement}, J.~C. {Raymond}, J.~{Vallerga}, M.~{Lemoine}, F.~{Dalaudier},
  J.~L. {Bertaux}, \aap {\bf 426}, 875--884 (2004)

\bibitem[{{Landgraf} et~al.(2000){Landgraf}, {Baggaley}, {Gr{\"u}n},
  {Kr{\"u}ger}, and {Linkert}}]{Landgraf+Baggaley+Grun+etal_2000}
M.~{Landgraf}, W.~J. {Baggaley}, E.~{Gr{\"u}n}, H.~{Kr{\"u}ger}, G.~{Linkert},
  \jgr {\bf 105}, 10,343--10,352 (2000)

\bibitem[{{Landgraf} et~al.(2003){Landgraf}, {Kr{\"u}ger}, {Altobelli}, and
  {Gr{\"u}n}}]{Landgraf+Kruger+Altobelli+Grun_2003}
M.~{Landgraf}, H.~{Kr{\"u}ger}, N.~{Altobelli}, E.~{Gr{\"u}n}, Journal of
  Geophysical Research (Space Physics) {\bf 108}, 8030 (2003)

\bibitem[{{Martin} et~al.(2007){Martin}, {Seibert}, {Neill}, {Schiminovich},
  {Forster}, {Rich}, {Welsh}, {Madore}, {Wheatley}, {Morrissey}, and
  {Barlow}}]{Martin+Seibert+Neill+etal_2007}
D.~C. {Martin}, M.~{Seibert}, J.~D. {Neill}, D.~{Schiminovich}, K.~{Forster},
  R.~M. {Rich}, B.~Y. {Welsh}, B.~F. {Madore}, J.~M. {Wheatley},
  P.~{Morrissey}, T.~A. {Barlow}, \nature {\bf 448}, 780--783 (2007)

\bibitem[{{Mathis}(1990)}]{Mathis_1990}
J.~S. {Mathis}, \araa {\bf 28}, 37--70 (1990)

\bibitem[{{Mathis} et~al.(1977){Mathis}, {Rumpl}, and
  {Nordsieck}}]{Mathis+Rumpl+Nordsieck_1977}
J.~S. {Mathis}, W.~{Rumpl}, K.~H. {Nordsieck}, \apj {\bf 217}, 425--433 (1977)

\bibitem[{{M{\"o}bius} et~al.(2004){M{\"o}bius}, {Bzowski}, {Chalov}, {Fahr},
  {Gloeckler}, {Izmodenov}, {Kallenbach}, {Lallement}, {McMullin}, {Noda},
  {Oka}, {Pauluhn}, {Raymond}, {Ruci{\'n}ski}, {Skoug}, {Terasawa}, {Thompson},
  {Vallerga}, {von Steiger}, and {Witte}}]{Mobius+Bzowski+Chalov+etal_2004}
E.~{M{\"o}bius}, M.~{Bzowski}, S.~{Chalov}, H.-J. {Fahr}, G.~{Gloeckler},
  V.~{Izmodenov}, R.~{Kallenbach}, R.~{Lallement}, D.~{McMullin}, H.~{Noda},
  M.~{Oka}, A.~{Pauluhn}, J.~{Raymond}, D.~{Ruci{\'n}ski}, R.~{Skoug},
  T.~{Terasawa}, W.~{Thompson}, J.~{Vallerga}, R.~{von Steiger}, M.~{Witte},
  \aap {\bf 426}, 897--907 (2004)

\bibitem[{{Redfield} and {Linsky}(2008)}]{Redfield+Linsky_2008}
S.~{Redfield}, J.~L. {Linsky}, \apj {\bf 673}, 283--314 (2008)

\bibitem[{{Savage} and {Sembach}(1996)}]{Savage+Sembach_1996}
B.~D. {Savage}, K.~R. {Sembach}, \araa {\bf 34}, 279--330 (1996)

\bibitem[{{Slavin} and {Frisch}(2007)}]{Slavin+Frisch_2007}
J.~D. {Slavin}, P.~C. {Frisch}, Space Science Reviews {\bf 130}, 409--414
  (2007)

\bibitem[{{Weingartner} and
  {Draine}(2001{\natexlab{a}})}]{Weingartner+Draine_2001a}
J.~C. {Weingartner}, B.~T. {Draine}, \apj {\bf 548}, 296--309
  (2001{\natexlab{a}})

\bibitem[{{Weingartner} and
  {Draine}(2001{\natexlab{b}})}]{Weingartner+Draine_2001b}
J.~C. {Weingartner}, B.~T. {Draine}, \apj {\bf 553}, 581--594
  (2001{\natexlab{b}})

\bibitem[{{Zubko} et~al.(2004){Zubko}, {Dwek}, and
  {Arendt}}]{Zubko+Dwek+Arendt_2004}
V.~{Zubko}, E.~{Dwek}, R.~G. {Arendt}, \apjs {\bf 152}, 211--249 (2004)

\end{thebibliography}


\end{document}